\documentclass{elsart1p}
\usepackage{graphics}
\usepackage{amssymb}

\begin{document}

\begin{frontmatter}



\title{Measurement of the Open Charm Cross-Section in $\sqrt{s_{NN}}$ = 200 GeV Cu+Cu Collisions at STAR}


%\author{Stephen Baumgart, Yale University, for the STAR Collaboration}
\author{Stephen Baumgart for the STAR Collaboration}
\address{Department of Physics, Yale University, New Haven, CT 06520, USA}

\begin{abstract}
% Text of abstract
We present the mid-rapidity $(D^{0} + \overline{D^{0}})/2$ yield measured in $\sqrt{s_{NN}}$ = 200 GeV Cu+Cu collisions at RHIC via direct reconstruction through the $K\pi$ decay channel. A charm cross-section is reported and compared to theoretical predictions and to previous RHIC experimental results.
\end{abstract}

\end{frontmatter}

% main text
\section{Introduction}
Charm is predicted to be produced primarily in gluon fusion reactions in the early stages of a collision \cite{Charm_Probe}. If charm is indeed produced in initial hard (high $Q^{2}$) processes, one will observe that charm production scales with the number of binary (nucleon-nucleon) collisions. The charm production cross-section can be calculated by summing up Feynman diagrams at the "Fixed-Order plus Next-to-Leading-Log (FONLL)" level. The total charm cross-section per binary collision as predicted by FONLL for $\sqrt{s_{NN}}$ = 200 GeV is $\sigma_{c\overline{c}}^{NN} = 0.256^{+0.400}_{-0.146}mb$  \cite{FONLL}. 

Both the STAR and PHENIX experiments at RHIC observe binary scaling when comparing p+p to Au+Au (PHENIX) and d+Au to Au+Au (STAR); however, previously measured charm cross-sections in STAR and PHENIX differ above declared uncertainties \cite{PHENIX_AuAu_D0,PHENIX_pp_D0,STAR_dAu_D0, STAR_AuAu_D0}. Also, current FONLL calculations under-predict measured STAR charm cross-section values \cite{STAR_dAu_D0}. This may suggest that different physical processes are at play than are assumed by FONLL calculations. 

%Both the STAR and PHENIX experiments at RHIC observe binary scaling scaling when comparing p+p to Au+Au %(PHENIX) and d+Au to Au+Au (STAR); however, previous STAR and PHENIX results are inconsistent with each %other \cite{PHENIX_AuAu_D0,PHENIX_pp_D0,STAR_dAu_D0}. Also results from the STAR experiment are higher %than FONLL predictions \cite{STAR_dAu_D0}. This may suggest that different physical processes are at %play than are predicted by FONLL calculations. 

\section{Experimental Setup and Analysis}
%This analysis uses data taken from the Time Projection Chamber (TPC) and the Time of Flight (TOF) detectors %at the Solenoidal Tracker at the RHIC (STAR), one of the detectors at the Relativistic Heavy Ion Collider %(RHIC).
This analysis uses data taken with the Time Projection Chamber (TPC) at the Solenoidal Tracker at RHIC (STAR), one of the experiments at the Relativistic Heavy Ion Collider (RHIC) at Brookhaven National Lab (BNL). The TPC measures an energy loss per unit distance, dE/dx, by particles' ionization of the P10 gas in the detector's chamber. The momentum is measured through the particles' curvature in a 0.5 T magnetic field \cite{TPC}. By using a Bichsel parameterization, a particle identification (PID) is acquired. 
%\subsection{The Time of Flight Detector}
%The Time of Flight Detector (TOF) patch is used in this analysis as a calibration device for the dE/dx %Bethe-Bloch bands in the TPC for a $p_{t}$ (transverse momentum) range of up to 1.8 GeV/c. The Time of %Flight Detector uses Multi-gap Resistive Plate Chamber (MRPC) technology. \cite{TOF_cite} By measuring %particles' velocities in the TOF in conjunction with momenta measured from the TPC, particle %identification (PID) can be achieved; however, its coverage is too limited to use it in $D^{0}$ %reconstructed.

The $D^{0} + \overline{D^{0}}$ invariant mass spectrum is reconstructed via the $K\pi$ decay channel. The $D^{0}$ (or $\overline{D^{0}}$) invariant mass is given by 
 \begin{equation}
  m_{D^{0}} = \sqrt{m^{2}_{\pi}+m^{2}_{K}+2(E_{\pi}E_{K}-|p_{\pi}||p_{K}|cos(\theta))}
 \end{equation}
where $\theta$ is the angle between $\overrightarrow p_{\pi}$ and $\overrightarrow p_{K}$.

\begin{figure}
\begin{center}
$\begin{array}{c@{\hspace{0.0in}}c}
\multicolumn{1}{l}{\mbox{\bf (a)}} &
	\multicolumn{1}{l}{\mbox{\bf (b)}} 
%\cr.\includegraphics{d0_cucu.eps} & \includegraphics{d0_cucu.eps}
\cr{\resizebox{7cm}{!}{\includegraphics{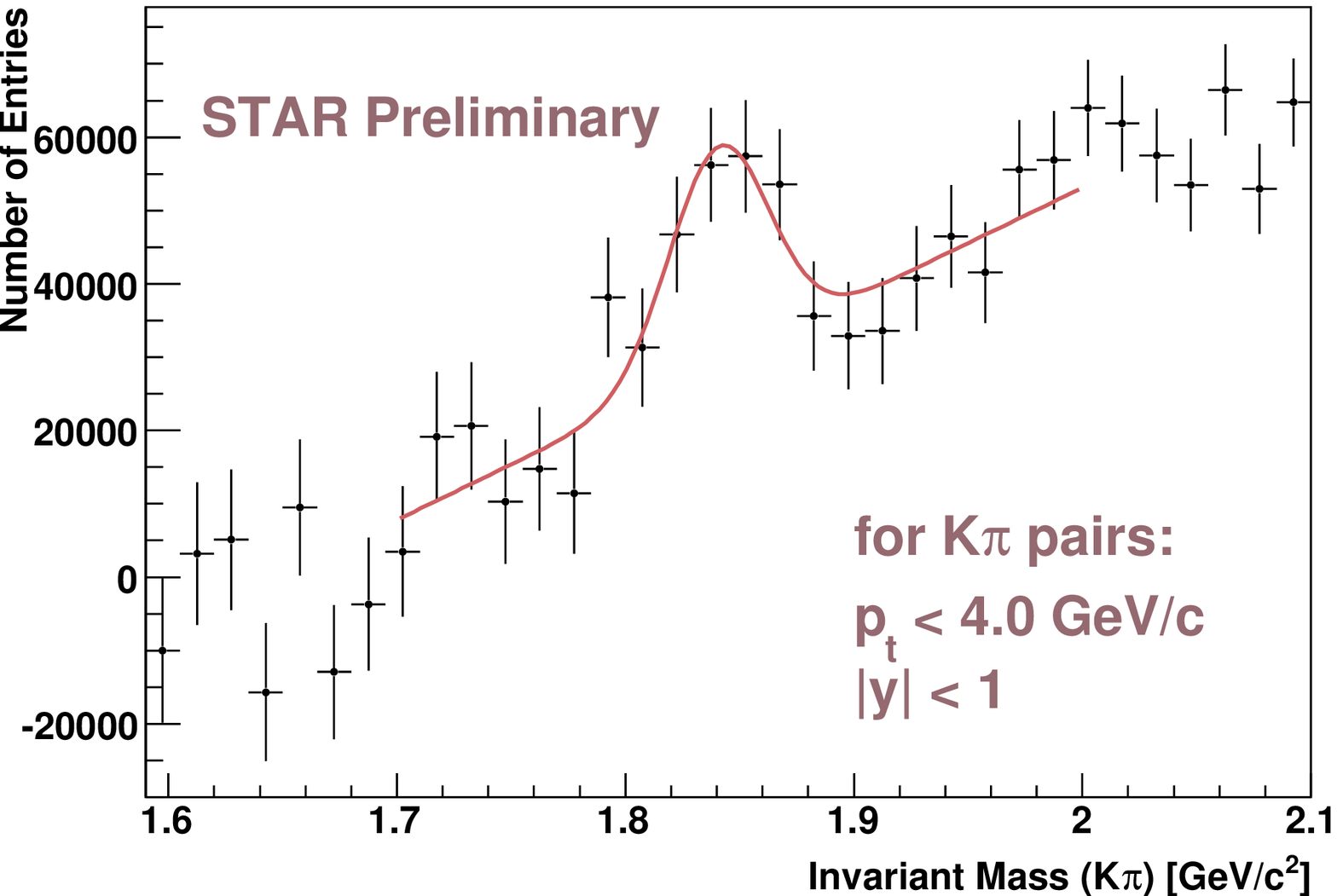}}}
 & {\resizebox{7cm}{!}{\includegraphics{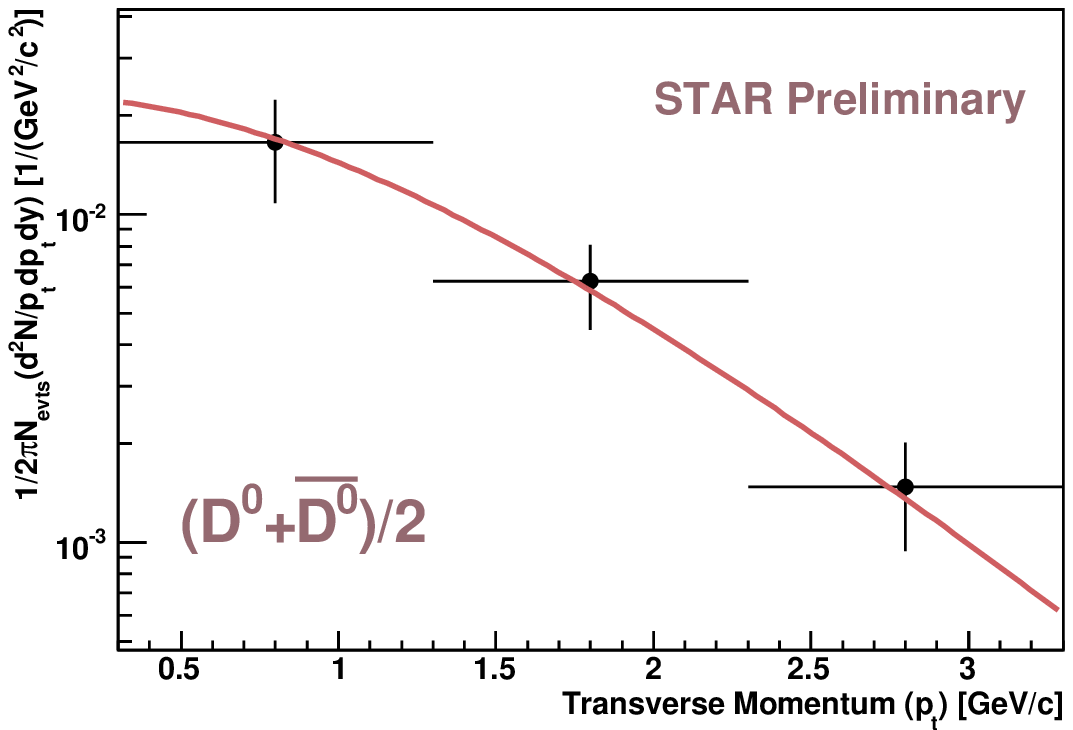}}}
%\epsfxsize=2in
%\epsffile{d0_cucu.eps} &
%	\epsfxsize=2in
%	\epsffile{d0_cucu.eps} \\ [0.4cm]
%\cr. \mbox{\bf (1)} & \mbox{\bf (2)}
\cr \mbox{\bf } & \mbox{\bf }
%\mbox{\bf (a)} & \mbox{\bf (b)}
\end{array}$
\end{center}
\caption{a) The $D^{0} +\overline{D^{0}}$ mass spectrum reconstructed from $K\pi$ in 28.7 million $\sqrt{s_{NN}} = 200$ GeV Cu+Cu minimum bias collisions after a rotational background subtraction. It is fit using a Gaussian function plus a first-order polynomial for the residual background. b) The $(D^{0} +\overline{D^{0}})/2$ transverse momentum spectrum fit with an $m_{t} - m_{0}$ exponential function.}
\label{D0}
\end{figure}

The invariant mass spectrum was extracted from 28.7 million Cu+Cu minimum bias events at $\sqrt{s_{NN}} = 200$ GeV. Then the uncorrelated background was subtracted via the rotational background subtraction method. To implement this method one of the daughter tracks in a two body decay is rotated in momentum space to every 5 degrees from 150 to 210 degrees in the plane transverse to the beam line. After the background is subtracted a peak with a statistical significance of $4.6 \sigma$ ($\sigma = \frac{S}{\sqrt{S+B}}$, S = area of signal peak, B = background under peak) is seen (See Figure \ref{D0}a). 

%\begin{figure}
%\begin{center}
%\rotatebox{0}{\resizebox{7cm}{!}{\includegraphics{d0_cucu.eps}}}\caption{The $D^{0} +\overline{D^{0}}$ %mass spectrum reconstructed from $K\pi$ in 28.7 million $\sqrt{s_{NN}} = 200$ GeV Cu+Cu minbias %collisions after a rotational background subtraction.}
%\label{D0}
%\end{center}
%\end{figure}

\section{Charm Cross Section}
\begin{figure}
 \begin{center}
\rotatebox{0}{\resizebox{7.5cm}{!}{\includegraphics{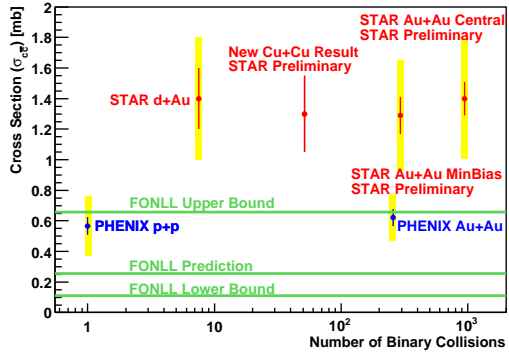}}}\caption{The measured charm cross-sections from STAR and PHENIX with the preliminary cross-section extracted for Cu+Cu as a function of the number of binary collisions. The green lines show the FONLL prediction and its upper and lower limits of uncertainty. 
%Consistency checks and a systematic error evaluation are still in progress.
}
\label{CuCu_Cross_Section}
 \end{center}
\end{figure}

Figure \ref{D0}b shows a transverse momentum spectrum for $(D^{0}+\overline{D^{0}})/2$ at mid-rapidity.  Efficiency and acceptance corrections, as calculated from a detailed GEANT simulation of the STAR detector, and a branching ratio correction of $3.80 \pm 0.07 \%$ \cite{Ratio} were applied. An $m_{t} - m_{0}$ exponential fit is done to the spectrum ($m_{t} = \sqrt{p_{t}^{2}+m_{0}^{2}}$). By integrating over the entire $p_{t}$ spectra, we find that $dN_{D^{0}}/dy$ = $0.184 \pm 0.035(stat.)$. The $dN_{D^{0}}/dy$ yield is converted to a cross-section via
 \begin{equation}
  \sigma_{c\overline{c}}^{NN} = (dN_{D^{0}}/dy)\times(\sigma_{pp}^{inelastic}/ N_{bin}^{CuCu})\times (f/R)
 \end{equation}
Here, $\sigma_{c\overline{c}}^{NN}$ is the charm cross-section per binary collision, $\sigma_{pp}^{inelastic} = 42$ mb is the proton+proton inelastic cross-section \cite{pp_cross_section}, $N_{bin}^{CuCu} = 51.5^{+1.0}_{- 2.9}$ is the average number of binary nucleon collisions in a Cu+Cu minimum bias collision found using the Glauber model, $f = 4.7\pm0.7$ is a normalization factor to extrapolate to full rapidity (based on Pythia simulation) \cite{STAR_dAu_D0, Pythia} , and $R = 0.54 \pm 0.05$ is the ratio of $D^{0}$ to $c\overline{c}$, measured in $e^{+}e^{-}$ collider events \cite{Ratio}.
The preliminary Cu+Cu charm cross-section is thus calculated to be  $1.30 \pm 0.25(stat.)$ mb and is shown in Figure \ref{CuCu_Cross_Section}. Evaluation of systematic error bars for the Cu+Cu measurement is in progress. This new measurement confirms previous STAR measurements \cite{STAR_dAu_D0,STAR_AuAu_D0} showing that charm production follows a binary scaling from p+p to central Au+Au collisions. 

\section{Conclusions}
STAR measured $D^{0} + \overline{D^{0}}$ in Cu+Cu collisions with a statistical significance of $4.6 \sigma$. The charm cross-section in $\sqrt{s_{NN}}$ = 200 GeV minimum bias Cu+Cu collisions was measured to be $1.30 \pm 0.25(stat.)$ mb. This is consistent with binary scaling when compared to other STAR measurements, verifying charm production via hard processes in the initial stages of the collision. 

\label{}

% \appendix

% \section{}
% \label{}

\end{document}